%% file: k7.tex
\documentclass[usenatbib,usegraphicx]{mn2e}
\usepackage{amsmath}
\usepackage{amssymb}
\usepackage{color}
\usepackage{graphicx}
\usepackage{graphics}
\usepackage{natbib}
\usepackage{enumerate}
\usepackage{hyperref}
\usepackage{times}
\usepackage{float}

\begin{document}
\bibliographystyle{mn2e}

\title[The KAT-7 Radio Telescope] {Engineering and Science Highlights of the KAT-7 Radio Telescope}
\date{Accepted 2016 April 29. Received 2016 April 28; in original form 2015 April 10}
\label{firstpage}
\pagerange{\pageref{firstpage}--\pageref{lastpage}}
\pubyear{2016}

\input{tex/authorlist}
\maketitle
\clearpage
\begin{abstract}
The construction of the KAT-7 array in the Karoo region of the
Northern Cape in South Africa was intended primarily as an
engineering prototype for technologies and techniques applicable to
the MeerKAT telescope. This paper looks at  the main
engineering and scientific highlights from this effort, and
discusses their applicability to both MeerKAT and other
next-generation radio telescopes.  In particular we found that the
composite dish surface works well, but it becomes complicated to
fabricate for a dish lacking circular symmetry; the Stirling cycle
cryogenic system with ion pump to achieve vacuum works but demands
much higher maintenance than an equivalent Gifford-McMahon cycle
system; the ROACH (Reconfigurable Open Architecture Computing 
Hardware)-based correlator with SPEAD (Streaming Protocol for 
Exchanging Astronomical Data) protocol data transfer
works very well and KATCP (Karoo Array Telescope Control Protocol)
control protocol has proven very flexible and convenient.
KAT-7 has also been used for scientific observations where it
has a niche in mapping low surface-brightness continuum sources, some extended HI
 halos and OH masers in star-forming regions. It can also be used to monitor
continuum source variability, observe pulsars, and make VLBI observations.
\end{abstract}

%\section{Introduction}
\input{tex/New_Introduction}
%\input{tex/abstract}

\input{tex/parameters}

%\section{Design Drivers}
%\input{tex/design}
%\section{Choice of Frequency}
\input{tex/frequency}
%\section{Antenna and Optics}
\input{tex/antenna}
%\section{RF and IF chain}
\input{tex/RF}
%\section{Correlator}
\input{tex/correlator}
%\section{Control}
\input{tex/control}

%%%%%%%%%%%%%%%%%%%%
\input{tex/8and9}
\input{tex/acknowledgements}
\bibliography{k7}
\vspace{5mm}
\input{tex/places}

%\bsp
\label{lastpage}
\end{document}

%% file: tex/authorlist.tex
\author[A.~R.~Foley et al.]
{A.~R.~Foley\thanks{E-mail:tony@ska.ac.za},$^1$ 
T.~Alberts,$^1$ 
R~P.~Armstrong,$^{1,2}$ 
A.~Barta,$^1$ 
E.~F.~Bauermeister,$^1$
H.~Bester,$^1$ 
\newauthor
S.~Blose,$^1$ 
R.~S.~Booth,$^1$
D.~H.~Botha,$^3$
% ...
\\
\textit{Affiliations are listed at the end of the paper}
}

%% file: tex/New_Introduction.tex
%\documentclass[usenatbib,usegraphicx]{mn2e}
%\usepackage{amsmath}
%\usepackage{amssymb}
%\usepackage{color}
%\usepackage{graphicx}
%\usepackage{graphics}
%\usepackage{natbib}
%\usepackage{enumerate}
%\usepackage{hyperref}
%\usepackage{times}
%%       \citestyle{aa}
%\usepackage{float}

%\hyphenation{HartRAO}
%\def\aap{Astronomy \& Astrophysics}
%\def\ApJ{The Astrophysical Journal}
%\def\AJ{The Astronomical Journal}
%\def\MNRAS{Monthly Notices of the Royal Astronomical Society}
%\begin{document}
%\maketitle

\section{Introduction}

\noindent

\begin{figure}
\includegraphics[width=\columnwidth]{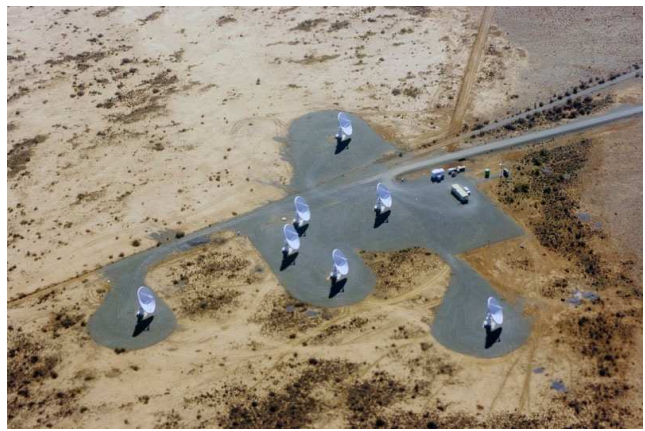}
\caption{An aerial view of the KAT-7 array; the antennas are pointing 
towards the south.}
\label{fig:aerial}
\end{figure}

The seven-dish Karoo Array Telescope array (KAT-7), shown in Figure~\ref{fig:aerial},
was built as an engineering testbed for the 64-dish Karoo Array Telescope, known as MeerKAT,
which is the South African pathfinder for the
Square  Kilometer  Array  (SKA).  KAT-7  and
MeerKAT are located close to the South African
SKA core site in the Northern Cape's Karoo desert
region about 80\,km north-west of Carnarvon. KAT-7 is remotely controlled from Cape
Town, some 800 km away from the site.

\begin{figure} 
  \includegraphics[width=0.99\columnwidth]{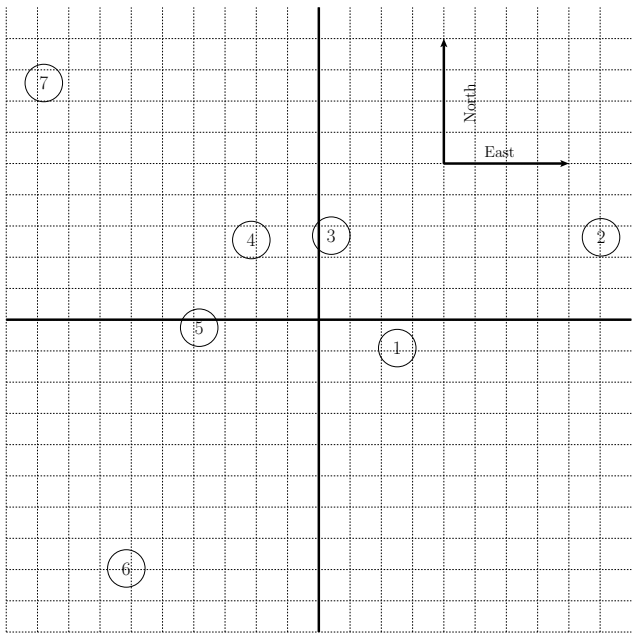}
  \caption{KAT-7, the 7-dish Karoo Array Telescope layout; each block
    is 10~m and the dish sizes are to scale. The zero point (where the
    thicker lines cross) is the array centre at $30.7148\degr$ S,
    $21.388\degr$ E.}
\label{fig:layout}
\end{figure}

The KAT-7 array is extremely compact, with baselines ranging between 26 m to 185 m.   The layout was determined using the optimization algorithm described in \citep{devilliers},
which determines a layout for a Gaussian UV distribution given a specified observation setting.
The observation setting being optimized in this
case was an 8 hour track, with a symmetric hour angle range, on a target at –60 degree declination.  The optimization objective was a Gaussian UV distribution at 1.4 GHz, yielding a Gaussian synthesized beam with low sidelobes.  Several randomly seeded layouts were generated and
were evaluated for a set of observing options ranging in 
time duration (snapshot, 4hr, 8hr, 12hr) and declination (0, $-30$, $-60$, $-90$ degrees).  The layout selected had the lowest sidelobes for the
largest number of test observation settings considered.  The antenna layout is presented in Figure~\ref{fig:layout}.

The KAT-7 dishes have a prime-focus alt-az design with a focal
ratio (F/D) of 0.38, each with a diameter of 12\,m, optimized for single-pixel
L-band feeds\footnote{\url{https://sites.google.com/a/ska.ac.za/public/kat-7}}.  The low noise amplifiers (LNAs) for the feeds are cryogenically cooled to 80~K using Stirling coolers. The key system specifications
for KAT-7 are summarized in Table~\ref{table:array}.

The digital backend of the KAT-7 system is an Field
Programmed Gate Array (FPGA)-based, flexible packetised correlator using the Reconfigurable Open
Architecture  Computing  Hardware  (ROACH
\footnote{\url{https://casper.berkeley.edu/wiki/ROACH}}), which
is a flexible and scalable system enabling spectral
line modes covering a wide range of resolutions.
Table~\ref{table:modes} gives the details of the recently commissioned correlator modes. Digital filters give a flat
bandpass over the inner 75\% of the band with a rapid roll-off at the edges of the band.

Originally, the Karoo Array Telescope was planned to consist of
20 antennas, and the main objective of building a radio telescope was
to support South Africa's bid to host the SKA.  The original design of the dishes were in fact driven by results from SKA optimization studies \citep{strawman}.  The scope of the project was later expanded to build the 64-antenna MeerKAT array.  Before building MeerKAT it was decided to build a smaller prototype array to field-test some of the
technologies that might be used in MeerKAT; that array is KAT-7. We note here that science goals for the 7-element array were considered secondary to the design and construction of the array elements. The original scientific considerations for the antenna layout and backend frequency coverage are considered in section~\ref{sec:freq}.

The new technologies implemented and tested on KAT-7 were expected to reduce either capital or running costs when compared to a more `traditional' radio telescope array. The novel parts included:

\begin{itemize}
\item On-site manufacture of single-piece reflectors made of a
  composite material using a vacuum infusion process. This had an
  embedded fine wire mesh to act as the radio reflective surface.
\item A single motor drive per axis with an anti-backlash mechanism
  and the ball-screw concept for elevation.
\item Stirling cycle cooling to cryogenic temperatures (80\,K) for the
  receiver.
\item Radio Frequency (RF) over fibre for data transport.
\item ROACH boards for all stages of correlation and beamforming.
\item Control using the Karoo Array Telescope Control Protocol (KATCP) (see section~\ref{sec:KATCP}).
\item Digital data transport in SPEAD (Streaming Protocol for
 Exchanging Astronomical Data) packets (see section~\ref{sec:SPEAD}).
\end{itemize}

The acquisition and construction phases of KAT-7 began in 2008 with the writing of the
telescope requirements specification. `First light' fringes, which
were the first successful observations by the interferometer, were obtained
between two antennas in 2009. This effectively marked the beginning of
the commissioning process. The commissioning and user verification process is complete, and KAT-7 is currently operational.

In addition to hosting the SKA and building MeerKAT, a further focus
area of the SKA South Africa project was to embark on a wide-ranging
Human Capital Development programme, to ensure that a new generation of
scientists and engineers would be available to use the MeerKAT and SKA
radio telescopes, and to further science and engineering in South
Africa in general.  To date many undergraduate and graduate students have been involved in the development of hardware and software for KAT-7 and MeerKAT.  Additionally, KAT-7 data has been used at several radio synthesis summer schools in South Africa and in many Honours, Masters, and PhD projects.

This paper is laid out as follows. In section~\ref{sec:freq} we discuss the reasons for the choice of frequency range for KAT-7. We then detail the antennas and optics (section~\ref{sec:ant}), the Radio Frequency and Intermediate Frequency (RF/IF) chain
(section~\ref{sec:RF}), the correlator (section~\ref{sec:corr}) and control systems (section~\ref{sec:control}). In sections~\ref{sec:images}--\ref{sec:comm} we describe some of the
early science and commissioning observations. We summarize technology lessons
learned and commissioning science results in section~\ref{sec:lessons}.

%\input{tex/parameters}

%\end{document}

%% file: tex/parameters.tex
\begin{table*}
\caption{Key performance parameters}
\begin{tabular}{lc}
\textbf{Parameters} & \textbf{Value} \\
\hline
Number of antennas        & 7 \\
Dish diameter             & 12\,m \\
Baselines                 & 26\,m to 185\,m \\
Frequency Range           & 1200\,MHz--1950\,MHz \\
Instantaneous Bandwidth   & 256\,MHz \\
Polarization              & Linear non-rotating (Horizontal + Vertical) feed \\
$T_{sys}$                 & \textless 35\,K across the entire frequency band \\
                         & ($\approx 30$\,K for all elevation angles \textgreater $30\degr$)\\
Antenna efficiency at L-band & 66\% \\
Primary beam FWHM at 1.8\,GHz & $1.0\degr$ \footnotemark[1] \\
Angular resolution at 1.8 GHz & 3\,arcmin \\
Location                 & latitude $30.7148\degr$ S, longitude $21.388\degr$ E, Elevation 1054m \\
Continuum Sensitivity    & 1.5\,mJy in 1 minute (256\,MHz bandwidth, $1\sigma$)\\
Angular Scales           & 3$\arcmin$ to 22$\arcmin$ \\
%\enddata
\hline
\end{tabular}
\\
$^1$The primary beam FWHM $\theta = 1.27\frac{\lambda}{D}$.\\
\label{table:array}
\medskip
\end{table*}
%table 2

\begin{table*}
\caption{Correlator Modes}
\begin{tabular}{lcc}
\textbf{Mode} & \textbf{Processed bandwidth/MHz} & \textbf{Channel bandwidth/kHz}  \\
\hline
Wideband   & 400 & 390.625\footnotemark[1]\\
Beamformer & 400 & 390.625\footnotemark[2]\\
8k Wideband & 400 & 48.8\footnotemark[3]\\
HI Galaxy Clusters & 400 / 16 = 25  & 25000 / 4096 = 6.1  \\
HI Large Galaxies  & 400 / 32 = 12.5 & 12500 / 4096= 3.0517 \\
HI Galaxies / Maser Search  & 400 / 64 = 6.25\footnotemark[4]  &  6250 / 4096 = 1.525879 \\
Maser Monitoring & 400 / 256 = 1.5625 & 1562.5 / 4096 = 0.3814697\\
\hline

\end{tabular}
\\
$^1$The channel bandwidth is obtained by dividing the IF  bandwidth (400 MHz) by the total number of channels (1024).  Note  that only 256 MHz of the IF bandwidth is usable due to RF filtering.\\
$^2$Similar to wideband mode, single boresight beam only.\\
$^3$The IF bandwidth (400 MHz) is divided into 8192 channels.\\
$^4$The usable Bandwidth is slightly less\\
\label{table:modes}
\end{table*}

%% file: tex/frequency.tex
\section{Choice of Frequency}
\label{sec:freq}
%based on stuff by Adriaan

\noindent
One particular scientific niche that was identified for a small-scale
interferometer in the southern hemisphere was to make high-sensitivity
observations of low surface-brightness emission from extended neutral
hydrogen in the nearby Sculptor cluster of galaxies. The
interferometer's response at 1.4\,GHz needed to be sufficient to
resolve structures on angular scales up to 24\arcmin\ (one third of
the antenna primary beam). This set the shortest baseline length at
30\,m. The positions of the six remaining antennas were selected to
optimize the interferometer response for 4-hour observations,
resulting in a randomized distribution with longest baseline of
180\,m.

The receiver frequency range was chosen to be 1200--1950\,MHz in order
to avoid potential interference from terrestrial Global System for Mobile Communications (GSM) and
aeronautical bands at 960--1164\,MHz. The 1.63:1 bandwidth ratio was still feasible with
corrugated horn feeds. This frequency range also gave the possibility
of joining in with standard 18-cm Very Long Baseline Interferometry (VLBI) observations of OH masers and continuum sources.

%% file: tex/antenna.tex
\section{Antenna and Optics}
\label{sec:ant}

\noindent
To confirm the suitability of the composite antennas for MeerKAT and the SKA
it was necessary to show that they had good aperture efficiency at the
highest frequency being considered, and that this lightweight structure
could be controlled in standard observing conditions.

To achieve aperture efficiency of 50\% at 10\,GHz with standard
horn feeds \citep{olver}, the unweighted
small-scale reflector accuracy was required to be better than 1.5-mm
r.m.s. This was achieved by combining theodolite
measurements of the completed antennas with finite-element analysis
for the worst-case load conditions. Follow-up measurements with radio
holography \citep{scott} using a number of geostationary satellite
beacons around 11.7\,GHz confirmed these results.

The antennas achieve blind pointing accuracy of 25\arcsec\ r.m.s.  under
all but the most extreme operating conditions, with a jitter of no
more than 5\arcsec\ over time scales of seconds. This is achieved at
tracking rates up to 0.05\degr/sec, across the elevation range. Below
2\,GHz the impact on measurements across the antenna beam is no more
than 1\%, and increases to a maximum of 5.5\% at 10\,GHz. Optical
pointing results show that referenced pointing \citep{rupen}, which
involves regularly checking the pointing against a pointing
calibration source near the field being observed, may allow accuracies
of better than 10\arcsec\ to be achieved, which would reduce the
impact to 2\%. The focal ratio was chosen to be 0.38, which is close to optimal for single pixel feeds using a conical horn feed \citep{rudge}.

\begin{figure}
\includegraphics[width=\columnwidth]{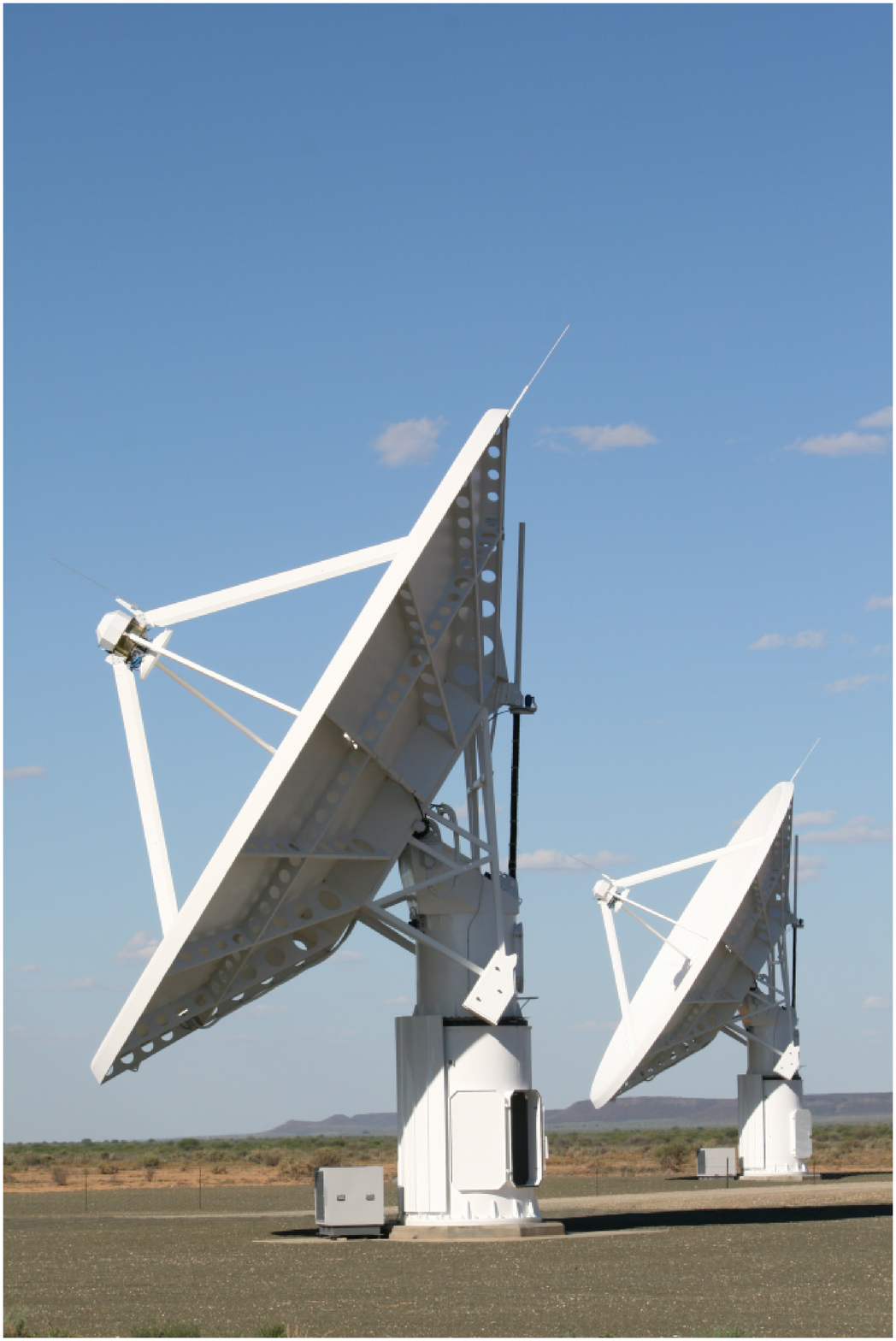}
\caption{Two of the KAT-7 dishes viewed side-on. Note the lightning rods at the apex and near the focus, the backing structure of steel beams with circular holes, the small counterweight on the right and the sun shield on the left hand side of the pedestal.}
\label{fig:CAD}
\end{figure}

\begin{table}
\caption{Dish specification}
\begin{tabular}{lc}
\textbf{Parameter} & \textbf{Value}\\
\hline
Pointing Accuracy & 25\arcsec \\
Surface Accuracy & 1.5~mm r.m.s.~(spec.)\\
 & 1.0~mm r.m.s.~(goal) \\
Specified Upper Frequency Limit & 10 GHz \\
Wind (Operational)  & 36 km/h \\
Wind (Marginal Operation) km/h & 45 km/h \\
Wind (Drive to Stow) & 55 km/h \\
Wind (Survival)  & 160 km/h \\
Azimuth Rotation slew speed  & 2\degr /s \\
Azimuth limits &  $-$175\degr, +285 \degr \\
Elevation slew speed & 1\degr /s \\
Elevation limits & 0\degr, 90\degr \\
Diameter &  12 m \\
Focal ratio f/D & 0.38 \\
Lowest Natural Frequency & 3 Hz \\
Feed/Cryo Mass  & 75 kg \\
Mount Type & Alt--Az Prime Focus \\
\end{tabular}
\label{table:ant}
\end{table}

A picture of two of the dishes is shown in Figure~\ref{fig:CAD} and dish
specifications are shown in table~\ref{table:ant}.
  Primary beam measurements at L-band were initially done by total
  intensity mapping using raster scans across very bright radio sources
  (in particular Hercules~A, Orion~A, Taurus~A and Virgo~A). These
  were found to be consistent with modeling done by Electromagnetic Software and Systems (EMSS), the company
  undertaking the receiver systems and optical design. Later the beam shape
  was measured more accurately using a full polarization
  holography-like \citep{scott} technique on the same bright sources (with some
  dishes tracking while others made spiral scans around the
  source). The broad features were the same, but the spiral scan mode
  showed details of the secondary lobes and the instrumental
  polarization in the primary beam.

\subsection{Why composites?}

 There are three main possible choices for a large dish antenna; a
 conventional steel structure based on panels, a lighter but possibly
 more expensive aluminium structure with panels, or a composite dish
 with metal backing for rigidity. For KAT-7 the choice was made for a
 composite for the following reasons:

\begin{itemize}
\item The dish front surface is constructed as a single unit, which
  gives it inherent stiffness (more efficient structure), something
  that is lost when the dish is constructed from loose panels.  This
  is an important aspect which reduces the mass while keeping the
  stiffness of the dish.
\item When constructed as a single unit there are no discontinuities
  in the reflective surface.
\item When constructed as a single unit there is no need to set up and
  align panels on a backing structure that is off the ground, which
  could prove to be time consuming and would be required for every
  dish.  Setting up of the mould is only required once, but must be
  done accurately.
\item Combining the composite dish surface with the steel rib and web
  backing gives a cost-effective solution.
\item An infusion process can be used, which is a tried-and-tested
  technology on such large structures.
\item Tooling is relatively inexpensive.
\end{itemize}

It is also very convenient that the thermal
  expansion coefficients of the composite dish front and the steel
  backing are very close to each other, giving low thermal
  loading. Thermal loading is potentially a large problem for pointing
  and surface accuracy, given the large temperature changes in the
  semi-desert climate of the Karoo. In addition the lightweight
nature of the dish meant that the counterweight needed would be light
and that the motors could be low-powered and still achieve good slew
speeds.  

The main drawbacks for a composite laminate
  dish are that the mould accuracy must be higher than that of the
  desired surface and that the layers of the laminate must be made in
  a balanced and symmetrical way to reduce the inherently anisotropic
  nature of the fibre mats. For balance any laminae must be placed
  symmetrically around a centre line, and for symmetry any laminae at
  layer (\textit{n}) must have an identical ply at layer
  (\textit{-n}) from the middle of the layers. For example a minimal
  set would be weave layers in the sequence (0,90)\degr\, (45,-45)\degr\, (45,-45)\degr\
  and (0,90)\degr\ . This symmetry is straightforward for a
  circularly symmetrical dish, but more complicated for other shapes, e.g. MeerKAT, which has an offset Gregorian geometry.

%% file: tex/RF.tex
\section{RF and IF chain}
\label{sec:RF}

\subsection{Feed}

\noindent
The feed provides native horizontal and vertical linear
polarization.  Conversion to circular polarization
in the receiver is not implemented into the design. Adding a polarizer would have increased the complexity of the receiver, which was considered a risk. Adding the required components (90\degr\ hybrid and isolators preceding the
low-noise amplifiers) inside the 80\,K cryostat with a single stage
Stirling cooler would increase system noise by at least 5\%.  A linear
polarization response could be calibrated sufficiently well to meet
the 30-dB linear and circular polarization purity required for
e.g.~measurements of Zeeman splitting; making a feed for a large fractional bandwidth (\textgreater~30\%) while retaining accurate circular polarization is difficult.

\subsection {RFE Architecture}

\begin{figure*}
\includegraphics[width=\textwidth]{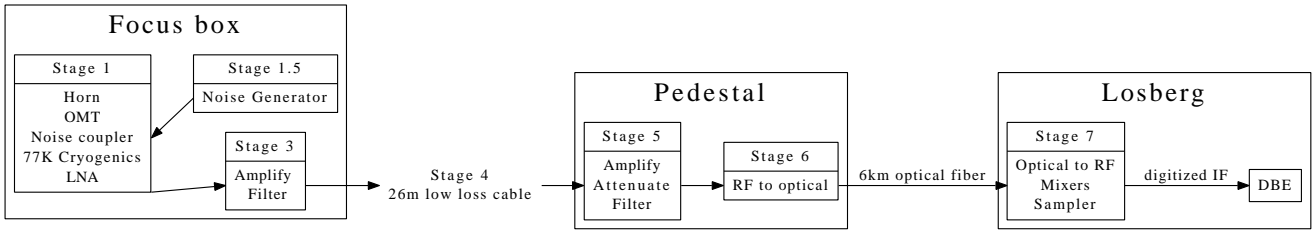}
\caption{Signal chain from receiver to Digital Back End (DBE). Signals
  arrive at the horn and go through the OrthoMode Transducer (OMT) to
  the Low Noise Amplifier (LNA). Calibration from a noise generator
  can be injected. After further amplification and filtering they
  descend to the pedestal where they are again amplified. The signal
  level is adjusted by attenuators and the signals is then filtered
  and used to modulate an optical signal. This is transmitted over
  optical fibre and converted back to an RF signal in the Losberg
  facility. This then is mixed with local oscillators and digitally
  sampled for the Digital Back End}
\label{fig:block}
\end{figure*}

\noindent
The Radio frequency Front End (RFE) architecture of KAT-7 is shown in 
Figure~\ref{fig:block}.
The design was driven by the following design decisions:

\begin{itemize}
\item Reduce complexity of the RFE at the antenna focus to reduce
  self-generated radio frequency interference (RFI) to the maximum extent possible, in preference to mitigating RFI later.
\item Reduce complexity of the RFE at the antenna location and move
  the down-converter and clock distribution to the same location as
  the digitiser.
\item Test the concept of mitigating RFI by moving the digitiser away from the antenna location to a shielded container located about 6\,km away, with the Losberg mountain providing a very high order of RFI shielding.
\end{itemize}

\begin{figure*}
\includegraphics[width=\textwidth]{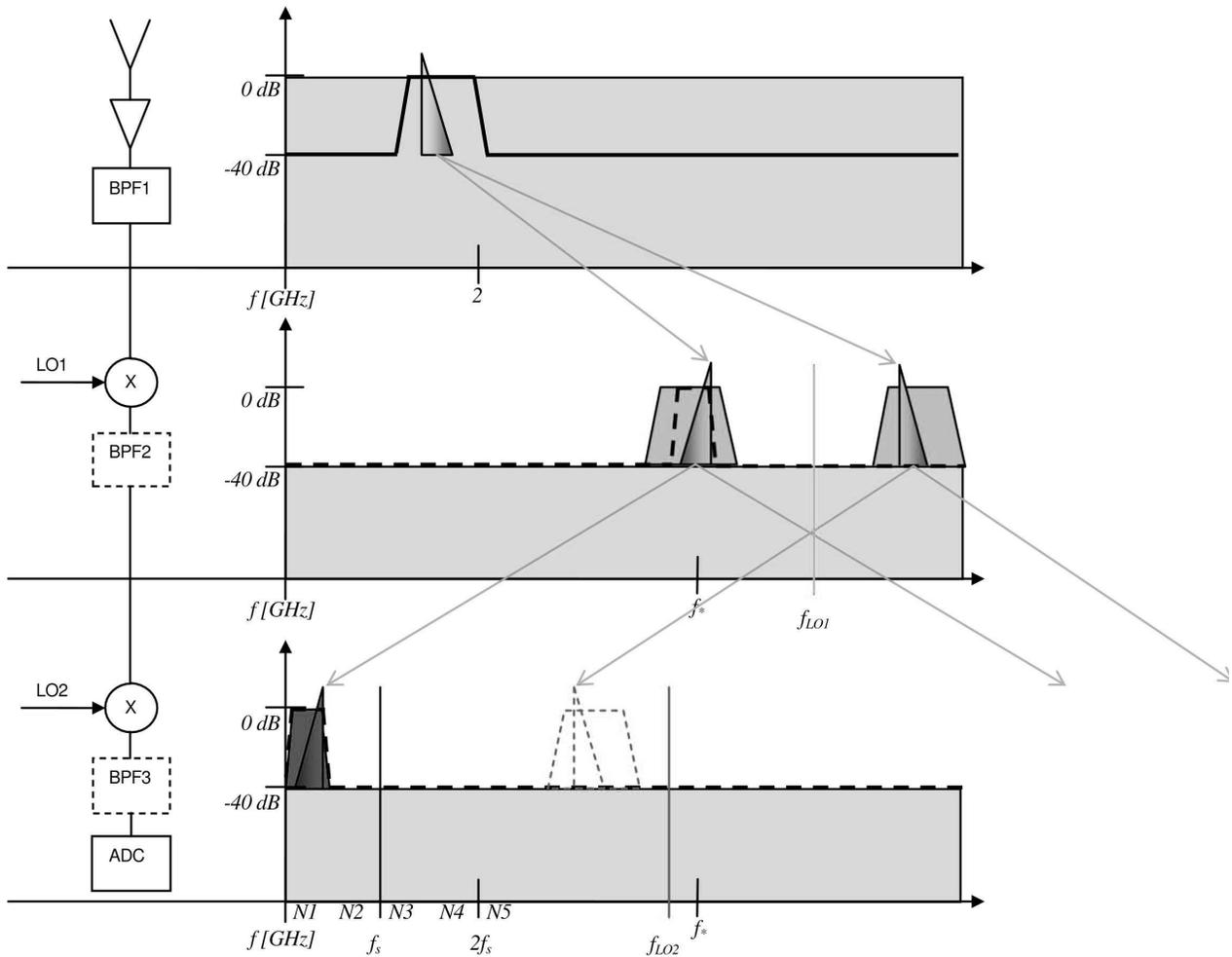}
\caption{An illustration of the down-conversion and sampling
  procedure. On the left side is the signal chain (top to bottom) with
  bandpass filters BF1-BF3, local oscillators LO1 and LO2 and the
  Analogue to Digital Converter (ADC). $f_{LO1}$ and $f_{LO2}$
  represent frequencies of the local oscillators. On the right are the
  spectra at the various stages. The top spectrum shows the signal as
  it arrives at Losberg, the middle spectrum shows the signal (and its
  alias) after mixing with LO1, while the bottom spectrum show the
  signal (and alias) after mixing with LO2, just before it is sampled
  by the ADC\@. N1 to N5 are the Nyquist zones. }
\label{fig:mixing}
\end{figure*}

\begin{figure*}
\includegraphics[width=\textwidth]{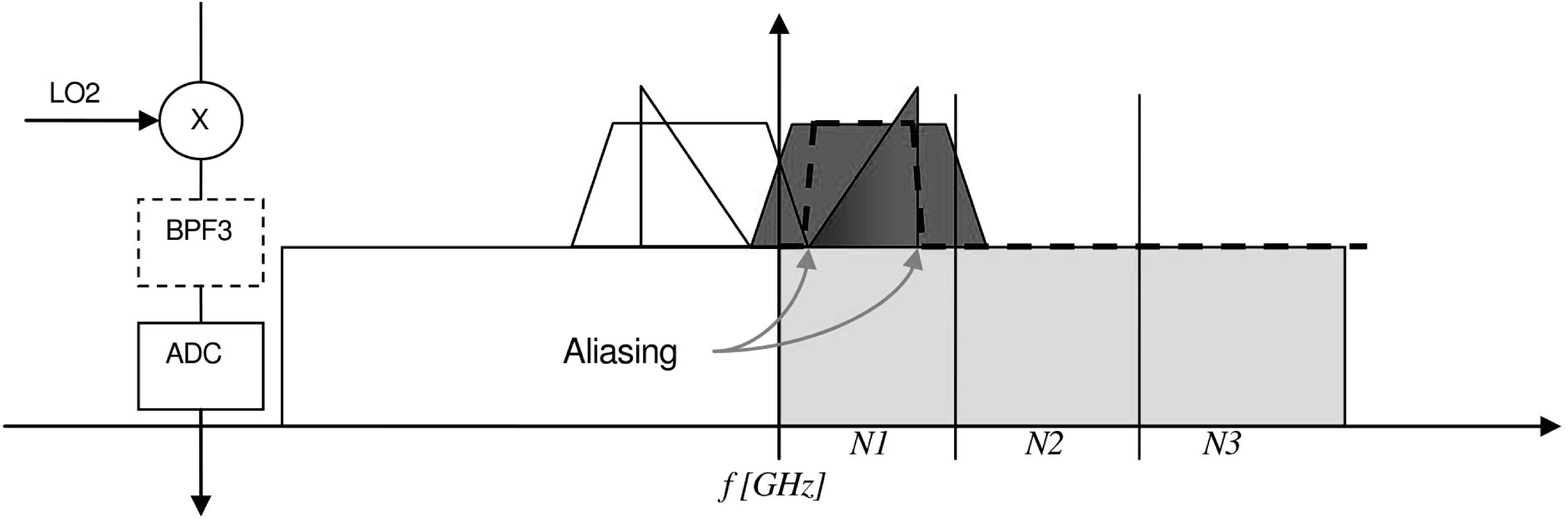}
\caption{Detail of the last stage of the down-conversion procedure,
  showing the last (fixed) local oscillator, LO2, and possible aliasing. The
  bandpass filter (BPF3) is designed to reduce that aliasing to a
  minimum before it enters the ADC, so that it does not appear in the
  sampled data. N1-N3 represent different Nyquist zones}
\label{fig:detail}
\end{figure*}

It is a double heterodyne system involving fixed and variable
local-oscillator (LO) signals that translate an RF (in the frequency
band of 1.2--1.95\,GHz) to an IF signal (in the band of 72--328\,MHz)
compatible with the digital samplers.  The first local oscillator
(LO1) can be set in the range 5.5--6.1\,GHz while the second (LO2) is
fixed at 4.0\,GHz.  The mixing scheme is shown in
Figure~\ref{fig:mixing}.  Figure~\ref{fig:detail} gives details of the
last stage of down-conversion and suppression of aliasing. `BPF's are
bandpass filters and `ADC's are analogue to digital converters.

\subsection{Noise Diodes}

%\textbf{XXXX SAY WHY WE HAVE A COUPLER *AND* A PIN XXXX}

\noindent
Each receiver includes two stabilized noise diodes to permit online
monitoring of system noise and receiver gain. The period and the duty
cycle of the noise system can be software selected, as well as having
a simple on/off mode. The signal can either be added to the received
(`sky') signal via an antenna that is integrated into the feed horn
(`the pin') or via 3-dB couplers that are installed between the
OrthoMode Transducer (OMT) ports and the low noise amplifiers (`the
coupler').  The `pin' signal level is approximately equivalent to the
system temperature (${T_{sys}}$) while the `coupler' level is
approximately $\frac{T_{sys}}{10}$. Using the `coupler' it is possible
to calibrate total-power measurements to an absolute scale with
accuracy better than 10\%.  The `pin' injection was designed to allow
delay and phase calibration between the orthogonal linear polarization
channels while the antennas were operating in single-dish mode, and
has been rarely used since the array was functional.  The receivers
were designed to permit observations of the Sun. However, no `high'
noise diode is provided to allow such observations to be calibrated.

\subsection{Down-conversion and Digitisation}

\noindent
To reduce the risk of RFI at the antennas, the analogue RF signal is
transported via fibre-optic cables to a central facility behind Losberg mountain. The signals are then
down-converted using a common LO, after which they are sampled in
baseband (i.e.~first Nyquist zone). The nominal signal is sampled with
3 bits (r.m.s.~$\approx 3$ levels) using 8-bit samplers. Together with
the 27\,dB of headroom to the 1-dB compression point provided by the
end-to-end analogue line-up, this provides the high dynamic range that
is required to limit the response to strong interfering signals. Even
on occasions when navigation satellites traverse the antenna beam,
gain compression remains at or below the 1\% level so intermodulation
products remain undetectable.

The analogue chain was designed to limit instrumental phase
instability to less than $8\degr$ r.m.s.~over 30-minute time scales. A Global Positioning System (GPS)-disciplined Rubidium maser clock is employed as frequency
reference to which the LOs are phase-locked. This allows the system to
achieve adequate phase stability to limit VLBI coherence loss to
20\% at frequencies up to 2\,GHz.

%% file: tex/correlator.tex
\section{Correlator}
\label{sec:corr}

\noindent
The correlator modes are noted in Table~\ref{table:modes}.  The
correlator is based on the so-called `FX' architecture (Fourier
transform `F' followed by Cross-correlation `X') with Hann-windowed
polyphase filter banks (PFBs).

The spectral-line modes are limited to a single band per observation,
which is a minor inconvenience for observations of OH maser lines. The
correlator always computes all four complex
  polarization products for all baselines (including
autocorrelations) since, with linearly-polarized feeds, total-power
measurements are easily corrupted by linear polarization and therefore
polarization calibration is required to achieve the highest possible
fidelity total-power imaging.

The correlator performs all internal computations with sufficient
numerical precision to support a spectral dynamic range of 43\,dB.

The Digital Back-End (DBE) also provides a coherently-summed
beamformer in parallel with a wide-bandwidth correlator. The
beamformer voltages are channelized to coincide with the correlator's
390\,kHz-wide channels. This correlator output is used by the Science
Processing Team's online pipeline to keep the
array phased-up during the course of the observation. The Science
Processor also provides the facility to re-synthesize and re-sample the
channelized voltages to the bandwidths required for the observation.
\begin{figure*}
\includegraphics[width=\textwidth]{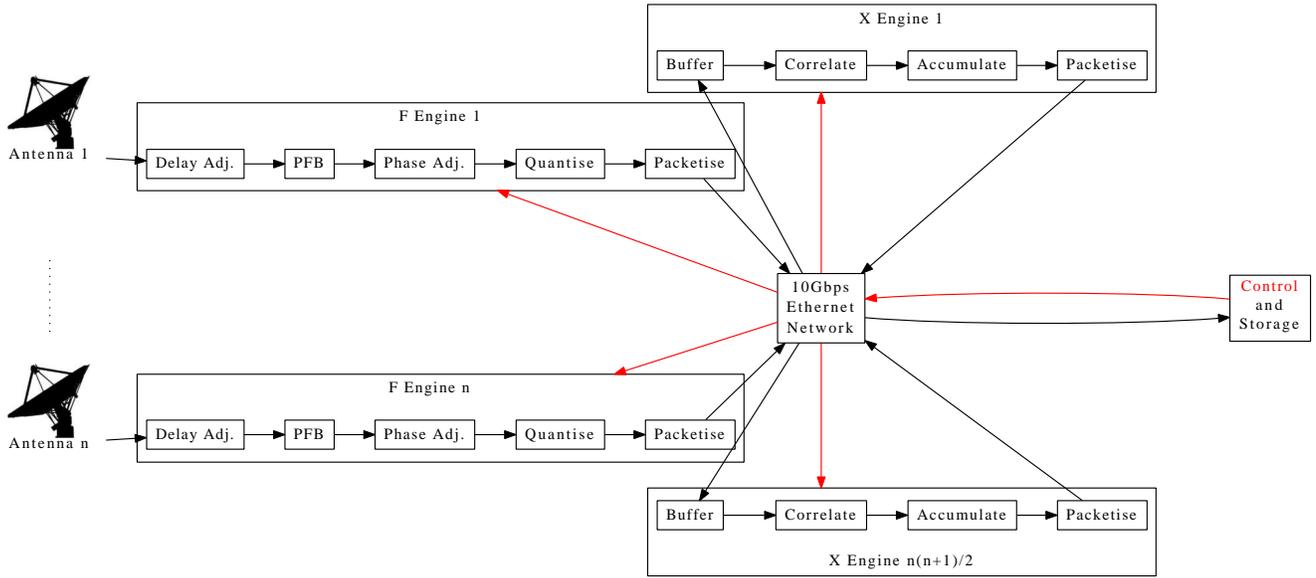}
\caption{The KAT-7 Correlator. The signal chain (in black) for two of
  the telescopes is shown, as well of the control and monitoring (in
  red). The F-engines do the Fourier transform and X-engines do the
  correlations. For $n$ antennas there are $\frac{n(n+1)}{2}$
  correlation products if you include all autocorrelations. For simplicity we consider only one polarization here.}
\label{fig:correlatorbd}
\end{figure*}
The beamformer's output may be used for VLBI recording or for
performing coherent de-dispersion of pulsar signals.
Figure~\ref{fig:correlatorbd} shows a simplified block diagram of
KAT-7's prototype correlator, consisting of three main components:

\begin{itemize}
\item The \textit{F-engines} which channelise the incoming data
  streams into spectral components.
\item The \textit{X-engines} which multiply and accumulate every
  product pair.
\item A commodity, off-the-shelf network switch to interconnect these
  boards.
\end{itemize}

KAT-7 required additional features to turn the basic correlator into a
more fully-featured digital back-end, most notably beamforming (for
tied-array operation) as well as additional operations in the
correlator itself such as fringe rotation and delay compensation, fast
readout speeds and a standardised output protocol. Hardware changes
were also required: The performance of the existing digitisers within
the Collaboration for Astronomy Signal Processing and Electronics Research (CASPER\footnote{\url{https://casper.berkeley.edu}}) collaboration
were found to be inadequate for KAT-7, and there was a desire for
KAT-7 to standardise on a single, general-purpose hardware processing
platform that could both digitise and process the signals.

In its current form, the KAT-7 DBE consists of 16 ROACH boards, eight of
which are configured as F-engines and the remaining eight as X-engines. It
is a real-time, full-Stokes, asynchronous packetized design with 16
inputs, processing 400\,MHz of RF bandwidth (only 256\,MHz of which
contains useful analogue signal because of the anti-aliasing filters
introduced in the RF chain (see Figure~\ref{fig:detail}).
ROACH boards can optionally host ADCs for digitising analogue signals
and the F-engines in KAT-7 are equipped with KAT ADCs to digitise the
down-converted IF signal at 800\,MSample/s with an 8-bit resolution. A
20-port 10\,GbE network switch is used to interconnect all the
processing boards.

While the F-engines and their ADCs are synchronously clocked from a
GPS-disciplined rubidium common clock source, the X-engines are purely
compute nodes that operate from their own, asynchronous,
clocks. Packets are timestamped in the F-engines and these headers are
interrogated by subsequent processors for data identification and
re-alignment. These X-engines could easily be
replaced by ordinary Central Processing
units (CPUs) and Graphical Processing Unit (GPUs) when technology advances enable
them to cope with the data rates presented.

The fringe rotation and delay compensation were implemented within the
F-engines by employing a combination of time-domain and
frequency-domain processing, and the system currently allows for
phased-tracked wideband (continuum) and spectral-line observations. It
does not allow online Doppler tracking.

Frequency-domain beamforming is implemented by tapping a copy of the
channelised data destined for the X-engines, and presenting it to a
co-located \textit{B-engine} which performs beam steering and
summation. The existing F-engines are thus shared by the X- and
B-engines. Development time and risk were reduced by leveraging the
existing delay-compensation software. This B-engine is not resource
hungry, needing only a single complex multiplier and adder to sum the
already serialised data, along with a lookup table for the steering
coefficients. This makes it possible to create additional beams with
modest incremental hardware costs.

All systems are being designed to be directly transportable and
up-scalable to larger arrays such as MeerKAT and the SKA.

%% file: tex/control.tex
\section{Control}
\label{sec:control}

\subsection{The KAT Control Protocol: KATCP}
\label{sec:KATCP}

\noindent
The MeerKAT telescope contains a large number of devices that require
control and monitoring, usually both. Most of these devices are
bespoke, meaning that they were constructed either within the project
or by a contractor to the specifications defined by the project.  As a
consequence we had the luxury of being able to select a communications
design strategy, architecture and protocol suite used for the control
and monitoring of the telescope.

A significant design decision was to confine low-latency, real-time
control functions to a small set of components. In other words the
control system as a whole does not operate on a low-latency basis ---
instead, time critical commands are sent out well in advance (with a time stamp in the future at which they are to be performed) and are
scheduled by a smaller real-time component close to the limited number
of elements requiring such control. The system distributes a clock
signal to subordinate nodes.  This decision makes it possible to use
computer systems running conventional (not hard-real-time) operating
systems that communicate using common, non-deterministic interconnects
and protocols. In particular it was possible to select transmission control protocol (TCP)/Internet protocol (IP) over
Ethernet for transport: these are well-established, pervasive and
inexpensive lower communications layers.

The only locally developed part of the control network is the
application control protocol, KATCP, running on top
of TCP/IP\@. KATCP borrows from established, classical Internet
protocols such as Simple Mail Transfer Protocol (SMTP), File Transfer Protocol (FTP) and Post Office Protocol (POP) in that it is an application protocol
which is easily read by humans; each line of text is a protocol
message. The protocol is not too different from a
command prompt or shell interface, where each input line is also a
command. However, unlike command line utility output, the protocol
replies have a regular, well-defined structure\footnote{See
\url{https://casper.berkeley.edu/wiki/KATCP}
 and \url{https://pythonhosted.org/katcp}.
}.

\subsection{Streaming Protocol for Exchanging Astronomical Data: SPEAD}

\noindent
In analogy with the control protocol, we unified the data exchange
protocol wherever practicable. Hence the correlator internals,
correlator output stage, and science processor all share a common
data-exchange format called
SPEAD\footnote{The specification document and a full description can
be found at \url{https://casper.berkeley.edu/wiki/SPEAD}}.

SPEAD is a one-way, best-effort, self-describing protocol, containing
both machine and human-readable descriptions. This allows receiver
processes to automatically unpack data, enabling application designers
to concern themselves with algorithms rather than with data
exchange. It also allows users to comment their data-streams so that
stored data can easily be reinterpreted at a later date without
additional documentation, or to easily debug data exchanges.

Other data protocols such as the VLBI Data Interchange Format (VDIF~\footnote{\url{http://www.vlbi.org/vdif/}}), were considered, but it was decided
that they were not sufficiently self-describing or flexible.

SPEAD was
developed in collaboration with the PAPER (Precision Array for Probing
the Epoch of Reionization; \citealt{parsons}) team as part of the
broader CASPER collaboration and is available under an open-source GNU general public license. SPEAD can be used for on-the-wire exchange, for on-disk
storage or for piping data between application processes within a single compute node.  This protocol is implemented on ROACH boards, CPUs, and GPUs.

\begin{figure}
\includegraphics[width=\columnwidth]{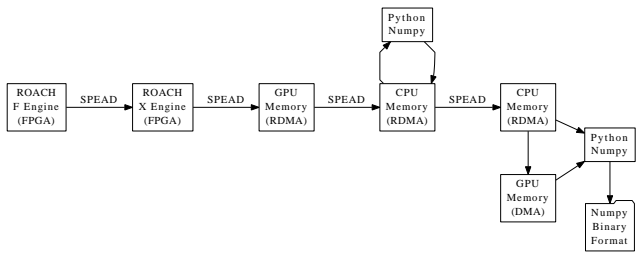}
\caption{SPEAD exchanges between a range of platforms. Once the data are sampled and Fourier transformed in the F-engine all the data are handled internally as SPEAD packets or as Numpy arrays.}
\label{fig:SPEAD}
\end{figure}

SPEAD is in essence optimised for the transmission of arrays of data
from one location to another. In particular we use SPEAD extensively
for the movement of 
\textsc{Python Numpy}~\footnote{\url{http://www.numpy.org}} arrays (Figure~\ref{fig:SPEAD}).  It
is designed to allow exchange of arbitrary data structures and aims to keep
receivers' copies fresh by propagating any changes
to the receiver as these changes occur on the transmitter.  The metadata and variable descriptors are injected into the primary data stream.
Such changes will occur if the correlator is started, or changes mode,
or the frequency of an LO is altered.
Receiver processes keep state information, so
the latest values of all variables within the structure are always
available, even if only the changed subset is transmitted.
If these dynamic features are not required, a lightweight and faster
static implementation can be made.

\label{sec:SPEAD}

As a one-way transport, and because it operates over the User Datagram Protocol (UDP) on Ethernet networks, it is easy to support multicast transparently, allowing
multiple devices to subscribe to the same data-streams.  This allows
real-time data inspection and plotting by subscribing to a subset of the 
data. As UDP does not check for packet delivery, SPEAD was made to tolerate
some packet loss, but cannot request a lost or corrupted packet.
Buffering helps by allowing for packets being received out of sequence.

%% file: tex/8and9.tex
\section{First Image with KAT-7}
\label{sec:images}

\begin{figure}
\includegraphics[width=\columnwidth]{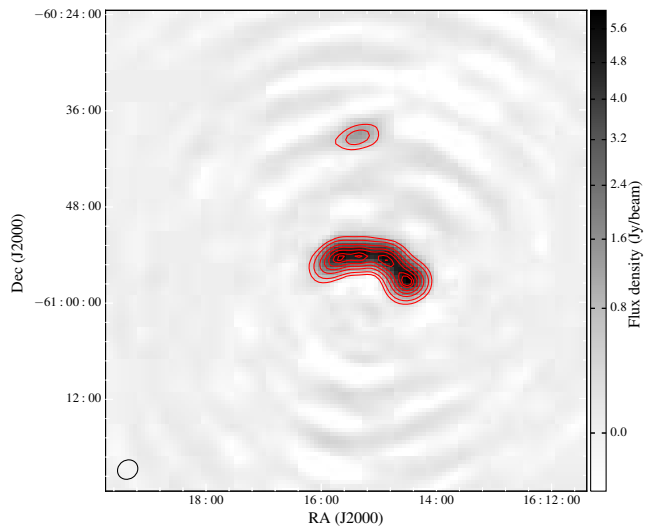}
\caption{Synthesis image of \mbox{PKS 1610-60} region containing the more
  prominent bent-double radio galaxy \mbox{PKS 1610-60.8} (below) and the
  smaller radio galaxy \mbox{PKS 1610-60.5} (above). Contours start at 0.5 Jy and increment by 1 Jy, with an additional contour at 1 Jy to better delineate \mbox{PKS 1610-60.5}. }
\label{fig:pks1610}
\end{figure}

\noindent
The first tests of the completed KAT-7 system were undertaken in 2011
August using the Wideband continuum mode (See Table~\ref{table:modes}). The specific combination of sensitivity, angular resolution, $\approx$ 3(4)$\arcmin$ at 1822(1328)\,MHz, and field-of-view ($\approx 1\degr$) of KAT-7 require extended southern sources  as imaging targets to use for testing this mode.  One such source is \mbox{PKS 1610-60} which has an extent of 30\arcmin\ in right ascension and
4\arcmin\ in declination \citep{christiansen}.  This source was
observed with KAT-7 for 10 hours with both polarizations at a
frequency of 1822\,MHz. The KAT-7 correlator outputs the data into an
observatory-specific format based on HDF5.  After conversion to a
standard Measurement Set the data are calibrated and imaged in CASA
\citep{mcmullin}. The final total-intensity image can be found in
Figure~\ref{fig:pks1610}. The dynamic range (ratio of the peak image intensity of 6.06 Jy to the r.m.s.~image noise  of 0.03Jy) of this map is about 200:1. We find a total integrated flux density of 30.9$\pm$3.1 Jy for \mbox{PKS 1610-60.8} and 1.6$\pm$0.2 Jy for the smaller radio galaxy \mbox{PKS 1610-60.5}. The measured fluxes are consistent with values listed in the Parkes Southern Radio Catalog (extrapolated from 1410 GHz assuming a spectral index $\alpha = 0.7$ where $S \propto \nu ^{-\alpha}$; \citet{wright}).

\section{Radio Continuum Imaging and Monitoring}
\label{sec:continuum}

\noindent
A sample of calibrators ($-90\degr<\delta<30\degr$ ) are continuously
being monitored and identified as possible short-spacings flux-density
calibrators \citep{kassaye}. The aim of this monitoring campaign is
also to image potential new calibrators with short spacings and to
assess their long-term variability. Several sources that are suitable
as point-source calibrators for interferometers with larger dishes,
and consequently smaller fields-of-view (such as e.g. \mbox{J0943-081}
and \mbox{J2326-4027}) have proven unsuitable for KAT-7 as they have
nearby contaminating sources within the primary beam.
%re-written -Tony
In general we have rejected candidate calibrators that have a contaminating source above 10\% of their peak brightness, or where the integrated flux density (estimated using the shortest baseline) is more than 10\% greater than the maximum flux density per beam.
%
% Most placed on the sky will have a few hundred milliJansky spread
% over a few sources within the KAT-7 primary beam at 1.2-1.95\,GHz.
Given the size of the KAT-7 primary beam at
  1.2 to 1.95\,GHz, most maps have few hundred milliJansky spread over a
  number of sources brighter than 1mJy/beam. This is consistent with source
  number counts from the FIRST \citep{white} and  ATLBS \citep{sub} surveys.  With the modest angular resolution of KAT-7 we would expect our images to be limited by confusion noise rather than thermal noise in Stokes \emph{I}. For most places on the sky we reach a continuum confusion limit of about 1(2) mJy/beam at 1822(1328)\,MHz
(based on no more than one source per five beams), and making high
dynamic range continuum images where confusion is so high has proven
difficult. The confusion limit is even higher along the Galactic plane.

From a full-polarization analysis, the polarization
  fraction of $8.4\pm0.6 \%$ was seen for 3C286 and $8.1\pm0.6 \%$ for
  3C138. These values are consistent with those determined by
  \citet{perley}.  Polarization was calibrated using full-Stokes
  observations of unpolarized radio sources such as \mbox{PKS
    1934-638} to determine instrumental leakage (of total intensity
  into polarization), and sources with high linear polarization at
  L-band over a large range of parallactic angles to measure delay and
  phase offsets between horizontal and vertical dipoles.  Typical
  leakages within the half-power beam width were of the order of 3\%
  in the upper band (1822\,MHz) and 3-6\% in the lower band
  (1322\,MHz), so second-order correction effects were negligible.

Although the dynamic range in continuum total
  intensity is limited by confusion after about 8 hours, this is not
  the case for polarization or spectral-line observations. However, the instrumental polarization corrections are needed for high
  dynamic range spectral line observations, such as in OH maser
  observations.

\section{Commissioning Science with KAT-7}
\label{sec:comm}

\noindent
KAT-7 went through a science commissioning phase.  The requirements
for commissioning science projects are (i) that they test the
performance of various aspects of commissioning (e.g.~correlator
modes, data reduction pipeline, RFI mitigation), and (ii) they
endeavour to contribute new and original scientific results.  These
projects touch on a broad range of scientific interest, including, but
not limited to, topics identified as MeerKAT Key Science Projects.
Projects include:

\begin{itemize}
\item Monitoring variable continuum sources (selected from the Astronomer's Telegram (ATEL) announcements).
\item Imaging extended, diffuse continuum emission.
\item Timing strong southern pulsars.
\item Conducting very long baseline interferometry (VLBI).
\item Measuring HI spectral line kinematics and low surface brightness emission in nearby galaxies and clusters.
\item Monitoring OH maser emission.
\item Mapping large scale polarization.
%\item Implementing techniques for HI intensity mapping.
\end{itemize}

\noindent
We outline several of these completed and ongoing science
commissioning projects in the following subsections.  We highlight
their scientific importance, and their contributions to KAT-7
commissioning and preparations for MeerKAT.

\subsection{Transient sources}

\noindent

Many transient sources
%(e.g.~PKS~-418, Circinus~X-1, PKS~2356-502)
have been monitored with KAT-7 ranging from neutron-star and black-hole X-ray binaries, blazars, Galactic gamma-ray binaries, and novae.  In this section we highlight several of these observations.

\begin{figure}
\includegraphics[width=\columnwidth]{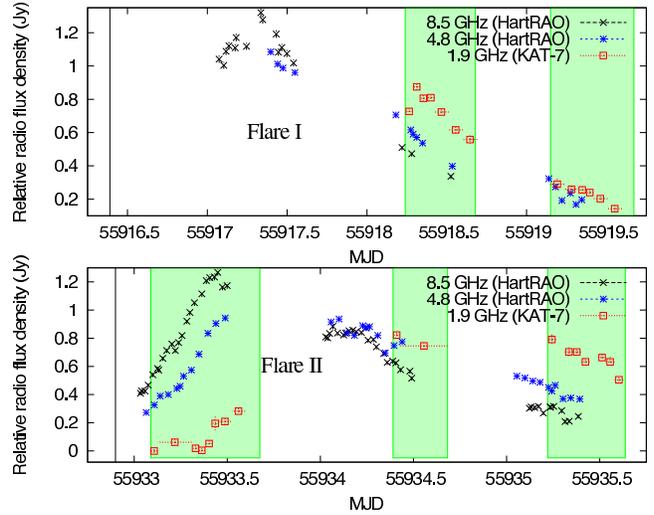}
\caption{A radio light curve of Circinus~X-1 obtained with KAT-7 (1.9\,GHz,
red crosses) and HartRAO (4.8\,GHz, black circles and 8.5\,GHz, blue stars). 
The X-ray binary system is clearly flaring to Jansky levels again. The times in which KAT-7 is operational are shown in green. The vertical bars at each
point represent the measurement noise, while the horizontal bars represent
the observation length.}
\label{fig:2F_OBS}
\end{figure}

Long-term monitoring of Circinus X-1, a bright and highly variable X-ray binary, was carried out during 2011 December 13 to 2012 January 19 using KAT-7 wideband continuum mode at 1.9\,GHz and the HartRAO radio telescope at 4.8 and 8.5\,GHz \citep{armstrong}.The observations confirm a return to strong radio flaring first observed in 1970's that had been suppressed for two decades. Figure~\ref{fig:2F_OBS} shows this long-term behaviour.  This is the first published scientific paper using KAT-7 data, and a good
example of multi-wavelength collaboration between South African instruments.

Simultaneous observations of the black hole candidate Swift
\mbox{J1745-26} with KAT-7, the Very Large Array (VLA), and the Australia
Telescope Compact Array (ATCA) radio interferometers reveal a `failed outburst'
event, illuminating details of the complex processes of accretion and
feedback in black holes \citep{curran}. The KAT-7 observations were carried out using the wideband continuum mode at a center frequency of 1.822 GHz over 13 epochs from 2012 September 13 to November 11. 

\begin{figure}
\includegraphics[width=\columnwidth]{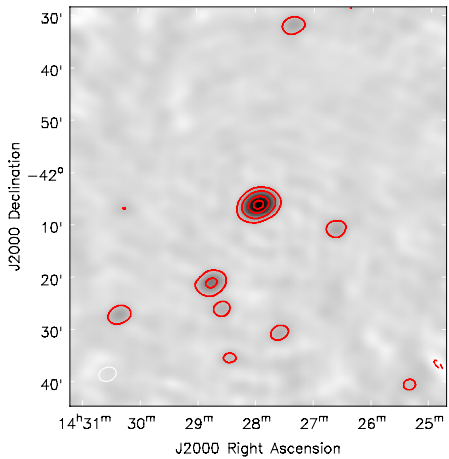}
\caption{Primary beam corrected image of \mbox{PKS 1424-418} imaged after ATEL4770.  Contours are -6, 6, 36, 180, and 360 times the r.m.s. noise of 8.5 mJy/beam. Negative contours are shown dashed and the synthesized beam is shown in the bottom left corner. The full width half maximum of the primary beam is 57.5 arcminutes.  All emission with a peak above 6 sigma of the r.m.s. noise are real non-variable radio sources and their fluxes are consistent with those extrapolated from a Sydney University Molonglo Sky Survey (SUMSS) image (assuming a typical spectral index of 0.7 for synchrotron emission).}  
\label{fig:atel4770}
\end{figure}

\mbox{PKS 1424-418}, a known blazar, was observed with KAT-7 as a follow-up to ATEL4770 \citep{Ciprini}.  At a redshift of $z=1.522$, this source has shown multiple flaring events in optical and $\gamma$ rays in 2009-2010 (\cite{Hauser2009},~\cite{Hauser2010},~\cite{Longo2009} and~\cite {Donato2010}). \cite{Tingay2003} showed that the source also shows variability in the radio regime. On 30th January 2013 it was reported in ATEL 4770 (“Swift detection of increased X-ray activity from gamma-ray flaring blazar PKS 1424-418”; \cite{Ciprini}) that PKS 1424-418 went through a $\gamma$-ray outburst and that the X-ray flux was increasing (as it had done this in the past).  This motivated us to obtain follow up observations with KAT-7 to search for variability in the radio wavelengths.  We observed PKS 1424-418 using the wideband continuum mode at a center frequency of 1822 MHz for 3.9 hrs on the 3 July 2013.  The primary beam corrected total intensity map is depicted in Figure~\ref{fig:atel4770}.  The r.m.s. noise in the image ranges from 8.5 mJy/beam near the phase center to 14.0 mJy/beam near the edge of the primary beam.  We measure a total  flux of 3.9$\pm$0.4 Jy for PKS 1424-418, consistent with the expected flux at the observed frequency for the target in its quiescent non-flaring phase~ \citep{Tingay2003}. All continuum emission located within the primary beam with a peak above 6$\sigma$ of the r.m.s. noise are real non-variable radio sources and their fluxes are consistent with those extrapolated from a Sydney University Molonglo Sky Survey (SUMSS) image (assuming a typical spectral index of 0.7 for synchrotron emission; \cite{Mauch}).  For instance for PKS 1425-421 (14h28m45.0s; -42d21m15s) we measure a total flux of 0.42$\pm$0.4 Jy and the extrapolated SUMSS flux at 1822 MHz is 0.44$\pm$0.01 Jy.

A campaign to monitor another another blazar source, \mbox{PKS 1510-089}, began on 21 October 2011 and lasted two years \citep{oozeer}.  PKS 1510-089 is a flat spectrum radio loud quasar at a redshift of $z=0.361$ and has been extensively observed from the radio to gamma wavelengths (\cite{Marsch}; ~\cite{Dammando1};~\cite{Dammando}; ~\cite{Saito}; ~\cite{Lazio}). The KAT-7 data were obtained using the wideband continuum mode at a center frequency of 1.822 GHz over 38 epochs with a cadence of at least once a month.  The integration time of each observation varied from 3 to 12 hours.  The r.m.s. noise of each observation ranged from 0.5 mJy beam$^{-1}$ to 2 mJy beam$^{-1}$.  At least one flaring event was successfully observed during this campaign (see Figure 2b of \citet{oozeer}).  The peak flux at quiescence is ~2.1 Jy and the peak flux at the peak of the flaring event is ~3.3 Jy.

\subsection{Diffuse radio emission}

\begin{figure}
\includegraphics[width=\columnwidth]{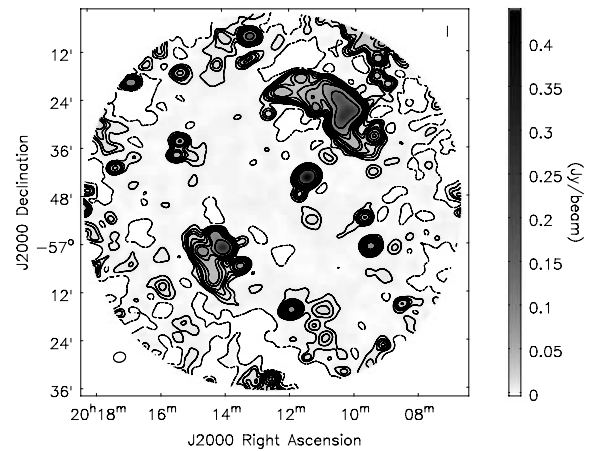}
\caption{Total intensity map of Abell 3667 at 1826\,MHz. Image noise
  is 1.30\,mJy/beam. Contours mark [-1,1,2,3,4,6,8,12,16,32,64,128]
  $\mathrm{\times \sigma_{rms}}$. The synthesised beam size (empty circle bottom left) is
  $\mathrm{184 \times 150 \arcsec}$.
  The FWHM of the primary beam is 1.0\degr.  The depicted region extends out to 10\% of the primary beam. Figure adapted from \citet{riseley}.
  }
\label{fig:aco}
\end{figure}

\noindent
With the availability of short spacings (26--185\,m, see Table~\ref{table:array}), KAT-7 is an ideal instrument to detect
extended diffuse radio emission especially in galaxy clusters hosting halos and relics. One such cluster is Abell 3667, one of the most well-studied galaxy clusters in the Southern Hemisphere.  It hosts two giant radio relics and a head-tail radio galaxy ~\citep{Rott}.  Recent work suggests that a bridge of diffuse synchrotron emission connects the north-western radio relic with the cluster center ~\citep{Carretti}.  To investigate the nature of the diffuse radio emission from the relics and `bridge' of the cluster we observed Abell~3667 with KAT-7 in March 2013.  The observations were done using the wideband continuum mode at a center frequency of 1826 MHz.  Nine closely packed pointings were used to cover the entire cluster region.  The mean time per pointing was 70 minutes.  The mean r.m.s. noise per pointing is 1.4 mJy beam$^{-1}$.  The pointings were mosaiced during deconvolution and imaged out to $10\%$ of the primary beam ($\simeq$ 1\degr at 1826 GHz).  As seen in Figure~\ref{fig:aco}, the nature of the diffuse emission (two large-scale relics) was confirmed, but only patchy low-level flux is observed between the central galaxy and the north-west relic.  This patchy emission is spatially coincident with several point sources that appear in the SUMSS catalogue \citep{Mauch}.  A more detailed analysis of these data can be found in \citet{riseley}.

\subsection{Pulsars}

\noindent
KAT-7 high time-resolution capability was first demonstrated through a $\simeq 2$~min single-dish observation of the Vela pulsar
(PSR~J0835$-$4510) in late 2012. These data were channelised into
$1024 \times 0.396$~MHz channels giving full 400~MHz bandwidth and
were sampled at the Nyquist rate of $2.56~\mu$s. The data were then
funnelled from the KAT-7 beamformer and written to disk using a
high-performance multiplexing algorithm to cope with data-throughput
requirements. A custom pipeline, based on {\small\textsc{dspsr}}
digital signal processing software
\citep{vb11}\footnote{\url{http://dspsr.sourceforge.net/}} was used to
produce total-intensity,
{\small\textsc{psrfits}}\footnote{\url{http://www.atnf.csiro.au/research/pulsar/index.html?n=Main.Psrfits}}
format data offline. {\small\textsc{psrchive}}, a scientific data
analysis software library was used for further processing and analysis
of these data
\citep{hotan}\footnote{\url{http://psrchive.sourceforge.net/}}.

Semi-regular dual-polarisation observations of bright pulsars
($\gtrsim 10$~mJy) using KAT-7 frequency-domain beamforming abilities
have since become standard practice. Phase-delay correction for
these observations is performed in a two-step procedure: 1)
coarse-delay correction through application of the known geometrical
delays between antennas and 2) fine-delay correction across the
band. The latter is performed through observation of a bright
calibrator source that is used to optimise the X/B-engine complex
weights. Beam-steering is subsequently performed in the X/B-engines to
produce a single, boresight beam per polarisation on the sky.

These observations rely on the same offline processing pipeline to
produce data in {\small\textsc{psrfits}} format, with at least 512
bins per pulse period, from the Nyquist-sampled, channelised voltage
data from the beamformer. Due to data-storage limitations, so-called
half-band mode is used where only the central 200~MHz of the observing
band (i.e. 512 frequency channels) is retained, as well as observing
time ranges typically between 5 and 15 min\footnote{An upcoming
  publication will discuss online de-dispersion and data processing
  and associated pulsar science with KAT-7.}.

\begin{figure}
  \centering
    %% Normal trim dimensions are left,bottom,right,top
    % NB Trim dimensions for 270 rot. are top, left, bottom, right
    \includegraphics[width=\columnwidth,height=8cm]{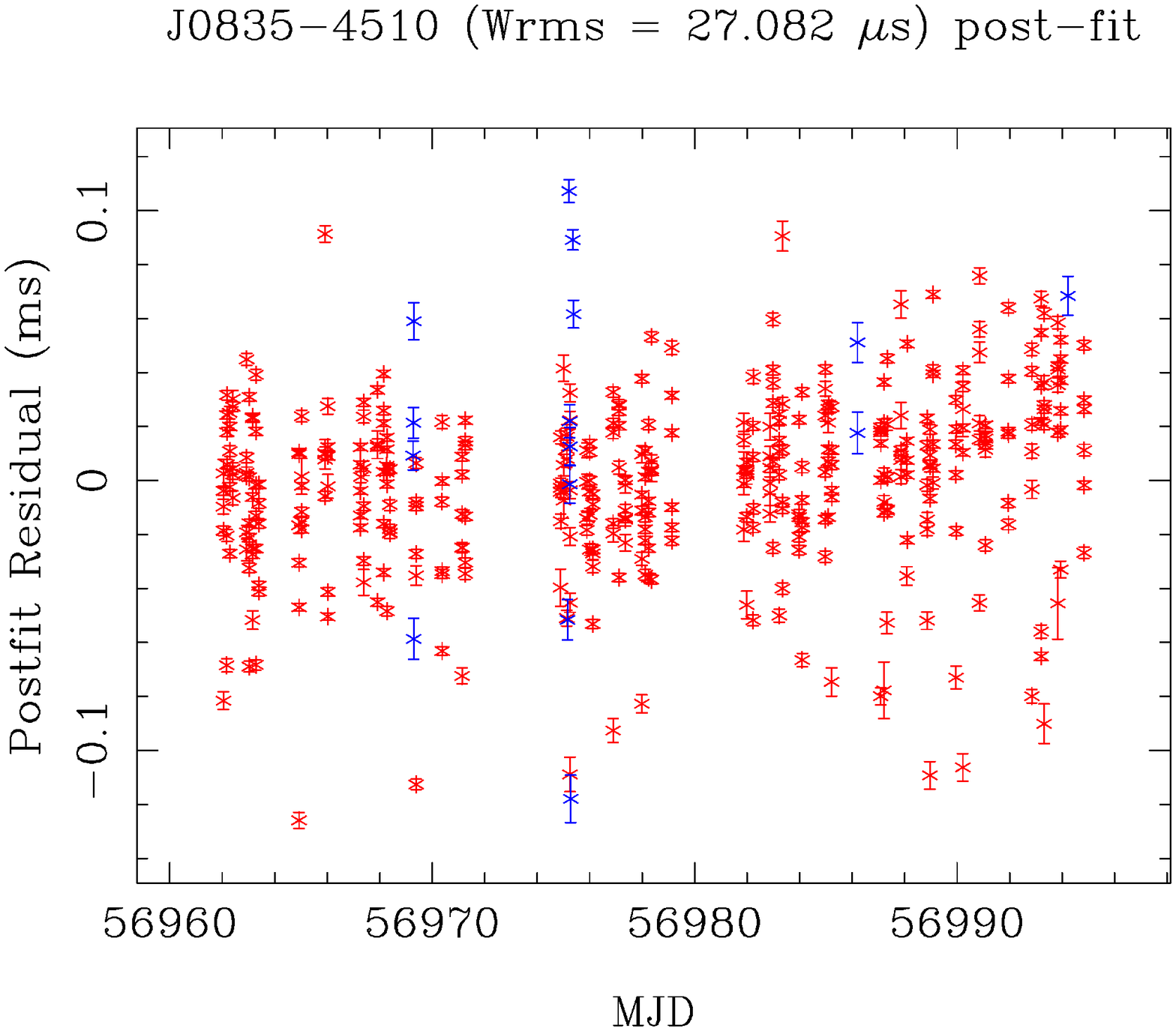}
\vspace{-4pt}\caption{
Timing residuals for the Vela pulsar (J0835$-$4510) over an
approximately 38 day span of data. Measurements obtained with the
HartRAO 26-m telescope (red crosses) were carried out almost daily and
typically lasted 1~min. These observations were centred on 1668~MHz with a bandwidth of 150~MHz. Those obtained with KAT-7 (blue
crosses) were obtained on 4 separate days only, with a typical
duration of 5~min. The KAT-7 observations were dual polarization with a bandwidth of 200~MHz and made use of all seven antennas. These data confirm that the timing accuracy of the
two telescopes is comparable.}
 \label{fig:0835res}
\end{figure}

\begin{figure*}
\centering
    %% Normal trim dimensions are left,bottom,right,top
    % NB Trim dimensions for 270 rot. are top, left, bottom, right
%\includegraphics[width=\columnwidth,height=8cm,angle=0]{figs/fig14.eps}
\includegraphics[scale=0.6,angle=0]{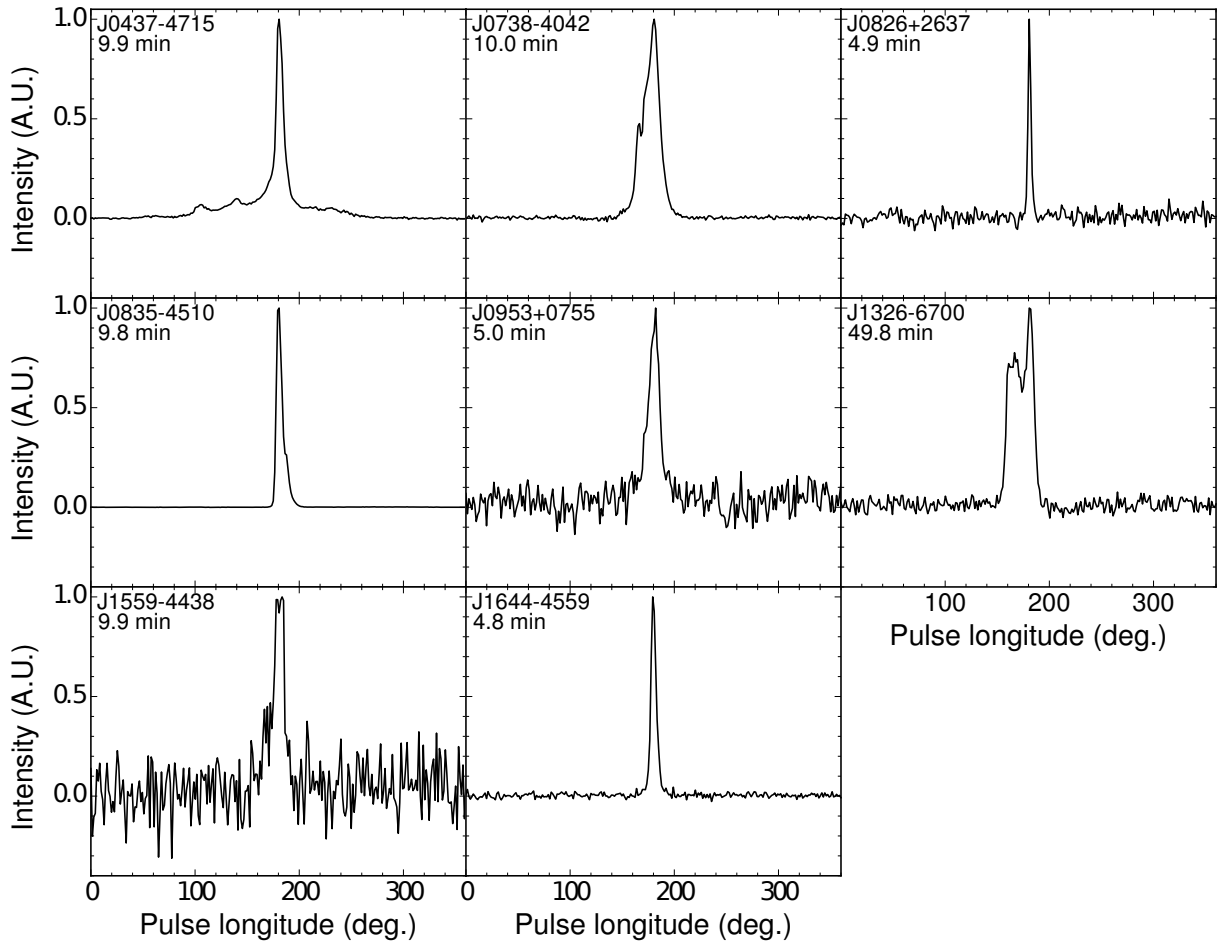}
\vspace{-4pt}
\caption{Average profiles for KAT-7 half-band observations of the
  pulsars J0437$-$4715, J0738$-$4042, J0826$+$2637, J0835$-$4510,
  J0953$+$0755, J1326$-$6700, J1559$-$4438 and J1644$-$4559. These
  observations were obtained between 2014~August~25 and 2015~April~24
  at a centre sky frequency of 1822~MHz. The observation durations are
  indicated in each of the respective subplots. The data were averaged
  across the frequency band and folded at the pulsar periods to
  produce average pulse profiles as a function of rotational phase.}
\label{fig:pulsar_profs}
\end{figure*}

Single-pulse and folded search capabilities have successfully been
demonstrated using a number of known sources. A series of short pulsar-timing
campaigns has also been undertaken, predominantly focusing on the
Vela pulsar due to its low period ($\sim89.3$~ms) and high flux
density (1.1~Jy at 1400~MHz; \citet{bf74}). Fig.~\ref{fig:0835res} shows a
comparison between the timing residuals obtained in 2014~November for
PSR~J0835$-$4510 using HartRAO and KAT-7 data. Fig.~\ref{fig:pulsar_profs} shows averaged
profiles for a number of pulsars produced via our custom data pipeline.

\subsection{VLBI}

\noindent
Pathfinder VLBI observations are critical to the establishment of the
African VLBI Network (AVN; \citet{gaylard}) in advance of SKA1.
Although KAT-7 is not equipped with a hydrogen maser, typically
required for VLBI observations, we obtained fringes between a single
KAT-7 antenna and the 26-metre antenna located at HartRAO near
Johannesburg, South Africa. The local GPS-disciplined rubidium time
source was deemed sufficient for this experiment. A short 10-second
test observation of the bright source 3C273 was performed on 2010
November 23. This involved HartRAO recording the data to their Mark~V
VLBI system, and KAT-7 recording to disk using the raw-voltage capture
system.

Due to the sampling clock of KAT-7 not being a power of 2, we used a
DDC (digital down-converter) to convert the sample rate in order to
produce the resultant 16~MHz bandwidth, via CUDA code deployed to a
GPU.
 
\begin{figure}
\includegraphics[width=\columnwidth]{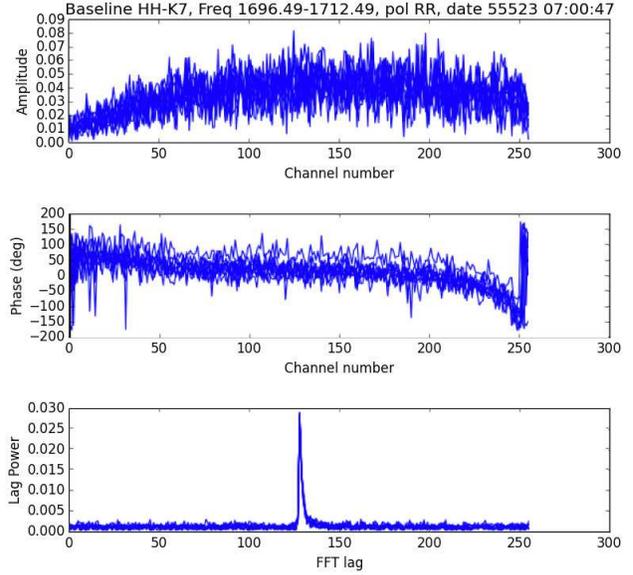} 
\caption{VLBI fringes obtained between KAT-7 and HartRAO using DiFX
  software correlator. Nine 1.024 second dumps are shown. The horizontal axis is frequency channel and
  the plots in descending order are visibility amplitude, phase (in degrees) and
  lag.}
\label{fig:vlbifringe}
\end{figure}

The data were combined and reduced using in-house python scripts to find the initial position and delay-offset values for the baseline. The
data were then combined using the DiFX correlator package with a integration time of 1.024 seconds and channelised with 256 channels
\citep{Deller}. Strong fringes were subsequently detected as shown in 
(Figure~\ref{fig:vlbifringe}).

\subsection{Spectral Lines: HI and OH observations}

\subsubsection{HI}

\begin{figure}
\includegraphics[width=\columnwidth]{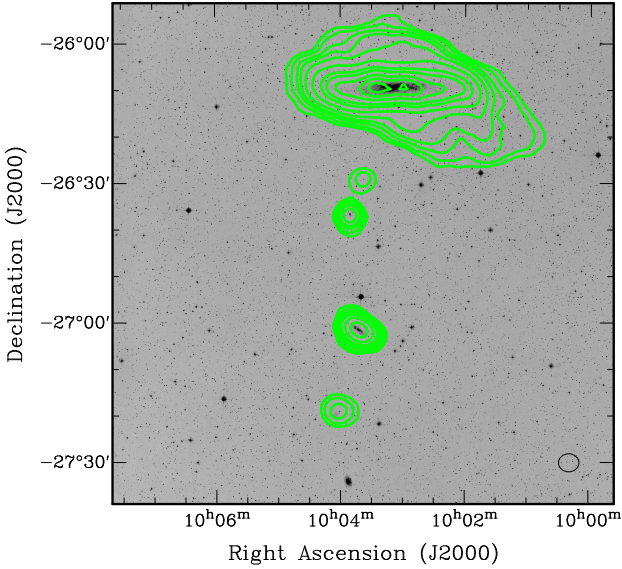}
\caption{HI mosaic of NGC~3109 and Antlia dwarf superposed on a
 DSS~B image. The green contours are [0.5, 1.07, 2.1, 3.2, 5.4,
 10.7, 21.4, 32.2, 53.6, 75.0, and 96.5] times the peak of 
 $107.2 \mathrm{Jy \,{beam}^{-1} \times km \,s^{-1}}$.
 NGC~3109 ($\mathrm{V_{sys}  = 404 \,km\, s^{-1}}$) is at the
 top and Antlia ($\mathrm{V_{sys} = 360 \,km\,s^{-1}}$) is at the bottom. The two
 background galaxies ESO~499-G037 ($\mathrm{V_{sys}  = 953\, km \,s^{-1}}$) and
 ESO~499-G038 ($\mathrm{V_{sys} = 871\, km \, s^{-1}}$) with its associated HI cloud
 ($\mathrm{V_{sys} = 912 \,km \,s^{-1}}$) are between NGC~3109 and Antlia. The
 synthesized beam is shown in the lower-right corner. Figure adapted from \citet{Carignan}}
\label{fig:ngc3109}
\end{figure}

\noindent
The receiver on KAT-7 covers the 1420\,MHz transition of neutral
hydrogen (HI) redshifted as far as 1200\,MHz (z=0.184),
although the intrinsic faintness of the
spectral line and the resolution of the telescope make it most
suitable to emission-line studies in the nearby universe. In
particular, KAT-7 is well suited for the study of nearby southern
objects which have emission on scales larger than 15$\arcmin$,
typically invisible to telescopes such as the VLA that lack the short
baselines. This was first illustrated in the KAT-7 HI observations of NGC~3109 \citep{Carignan}, where 40\% more HI flux was detected
than in previous VLA measurements (see Figure~\ref{fig:ngc3109}). Observations of NGC~3109 and the dwarf galaxy Antlia were carried out over 13 observing sessions between 2012 November 20 and 2012 December 26 using the `HI galaxies' correlator mode (see table \ref{table:modes}) while this was being commissioned.  Three fields were positioned in a straight line and extending slightly to the SE to mosaic the region between NGC~3109 and Antlia, covering roughly a region of 2.5\degr by 1\degr on the sky.  Each field was observed for 25 hours resulting in an r.m.s. noise of 3.7 mJy beam$^{-1}$ in a 1.28 km s$^{-1}$ channel. The new data also allowed the derivation of the rotation curve 16$\arcmin$ further out than the previous VLA measurements (see Figure 10 of
\citet{Carignan}). A detailed look at the data during
calibration led to the discovery of a source of antenna-dependent,
faint, very narrow, internally-generated RFI that was successfully
eliminated by the insertion of a low-pass filter along the signal path.

\begin{figure}
\includegraphics[width=\columnwidth]{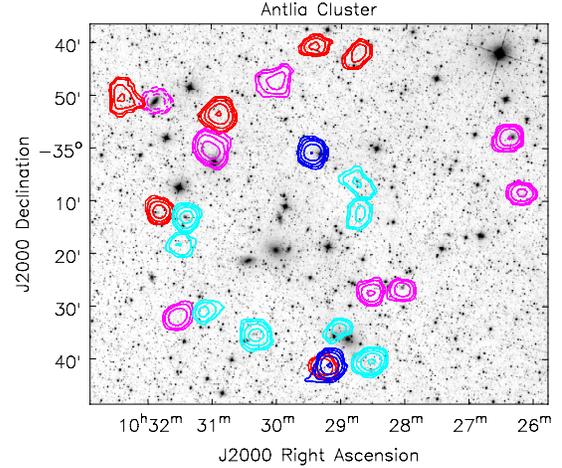}
\caption{A cut-out of the Antlia Cluster mosaic: color contours
  corresponding to the systemic velocity of the galaxy detected in HI
  emission are overlaid on a WISE 3.4\,$\mu$m image. NGC~3281,
  detected in absorption, is in dashed contours.  Dark blue is
  $\mathrm{<2200 \,km \, s^{-1}}$, cyan is $\mathrm{2200-2800 \,km
    \,s^{-1}}$, magenta is $\mathrm{2800-3400 \,km \,s^{-1}}$, and red
  is $\mathrm{>3400 \,km\, s^{-1}}$.  HI rich galaxies are detected in
  a ring around the dominant elliptical galaxy, NGC~3268. These KAT-7
  observations had a usable bandwidth sensitivity to HI emission at
  recessional velocities between $\mathrm{1200-4600 \,km \,s^{-1}}$,
  and 37 HI galaxies were detected between $\mathrm{1800-4300 \,km
    \,s^{-1}}$.  Figure adapted from \citet{hess}}
\label{fig:antlia2}
\end{figure}

The wide field-of-view of KAT-7 and wide-bandwidth spectral-line modes also
make it a powerful instrument for mosaicing large areas on the sky,
competitive with ATCA in terms of total observing time for similar
sensitivity.  As part of commissioning the 25\,MHz `HI Galaxy
Clusters' spectral-line correlator mode, several projects
to image the HI emission in galaxy groups and clusters are currently underway
(e.g.~Antlia Cluster, NGC~4055 Group). In particular, the Antlia
Cluster was observed for 147 hours in a 7-pointing discrete hexagonal
mosaic, covering roughly $4.4\mathrm{deg^2}$ , and reaching 
$0.97 \mathrm{mJy \,beam^{-1}}$
over the $15.5 \mathrm{km \,s^{-1}}$ channels (see Figure~\ref{fig:antlia2}).  This deep map detected 37 HI cluster members, 35 of
which were new HI detections, and 27 of which were the first redshift
measurements at any wavelength.  The broad bandwidth coverage, and
lack of bright, dominating sources in the Antlia mosaic led to the
recognition of the `$u=0$' problem in KAT-7 spectral-line data.  In
this case, RFI that exists below the level of the noise in the
visibility data, and would normally be incoherent, adds constructively
when the fringe rate between two antennas is equal to zero.  It is not
a unique phenomenon to KAT-7, but it is particularly a problem on
short baselines.  (See \citet{hess} for a more complete explanation).

\subsubsection{OH}

\noindent 
The frequency range of KAT-7 covers all four ground-state transitions
of hydroxyl: 1612, 1665, 1667 and 1720\,MHz.  Maser emission occurs in
these transitions under a variety of conditions: in massive
star-forming regions (primarily in the main-lines at 1665 and
1667\,MHz \citep{caswell98}, in the shells around AGB stars at
1612\,MHz \citep{sevenster}, and in shocked regions such as supernova
remnants at 1720\,MHz \citep{wardle}.

While KAT-7 does not have the angular resolution to map individual
maser spots, its narrowest spectral mode gives a velocity resolution
of 68\,m/s, enough to resolve narrow maser lines in massive
star-forming regions.  A number of methanol masers in star-forming
regions have been found to exhibit periodic variations
(\citet{goedhart} and references therein). \citet{green} attempted
to monitor a short-period source in the hydroxyl main-lines and found
a weak indication of periodicity, but the time-series was
undersampled. Thus monitoring is an ideal niche application for KAT-7 since
it generally has more time available than the Parkes Telescope or
ATCA\@. Six of the known periodic methanol masers have hydroxyl maser
counterparts that can be detected with KAT-7. These sources are
being monitored on a weekly basis at both 1665 and 1667\,MHz, using
interleaved observations at both frequencies in a 13-hour schedule
block.

\begin{figure*}
\includegraphics[width=\textwidth]{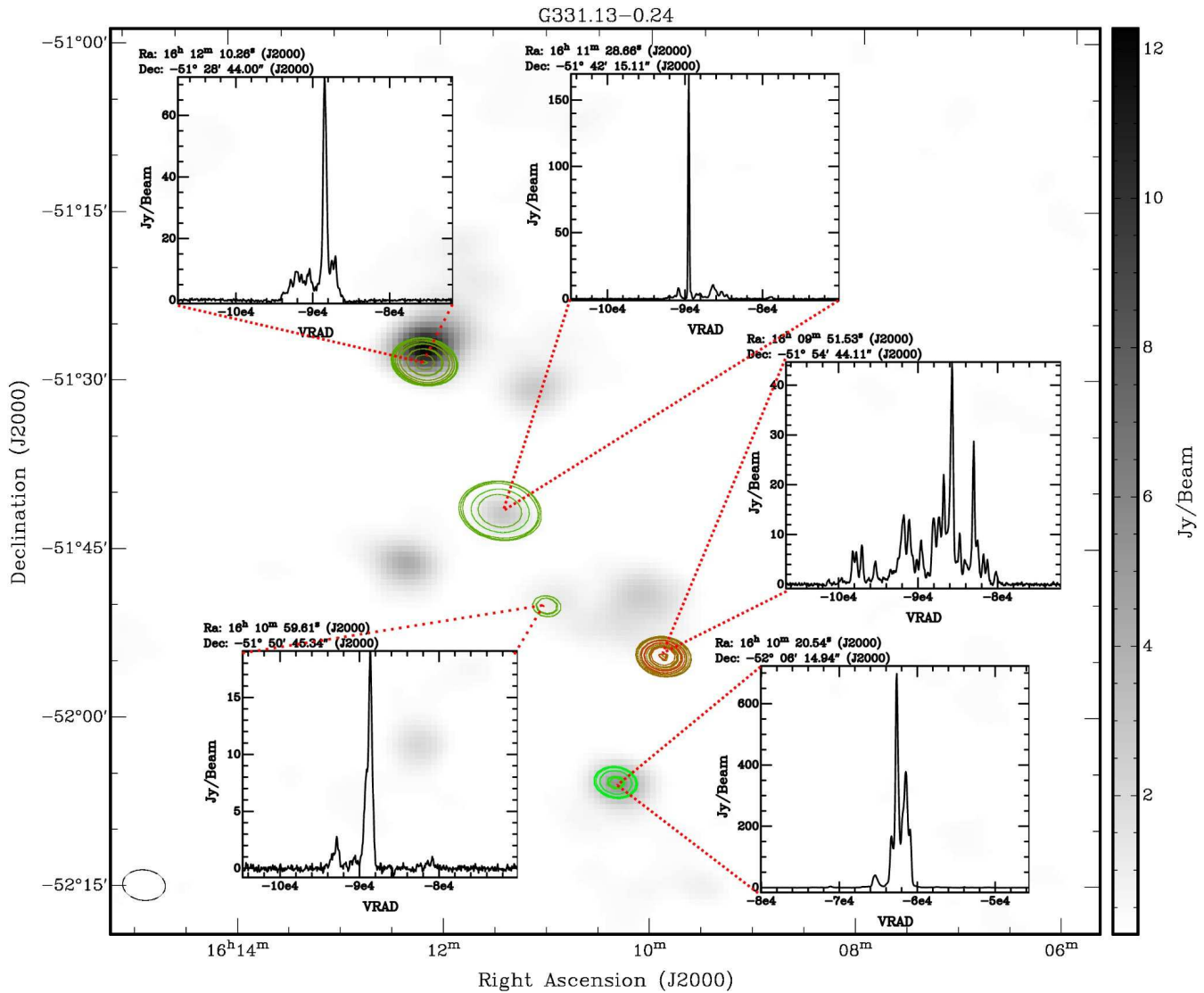}
\caption{Masers in \mbox{G331.13-0.24} at 1665\,MHz monitored with KAT-7. The
  underlying greyscale map shows the positions of the continuum
  emission while the coloured contours show the continuum-subtracted
  maser emission. The insets show the spectrum at the indicated
  positions, where the radial velocity scale is $\mathrm{m \, s^{-1}}$ with
  respect to
  the Local Standard of Rest. The synthesized beam is shown as a
  dashed ellipse at the bottom left corner. Primary beam correction was not applied.}
\label{fig:masers}
\end{figure*}

Figure~\ref{fig:masers} shows a pilot observation, carried out on 2013 February 22,  of the star-forming
region \mbox{G331.13-0.24} at 1665\,MHz. The large field-of-view of KAT-7 led
to the detection of several maser sources as well as several diffuse
HII regions. The positions, velocities and flux densities are
consistent with those from recent Parkes observations
\citep{caswell14}, confirming that the system is performing as
expected.  The observations are dynamic-range-limited and three of the
six fields observed (including \mbox{G331.13-0.24}) are dominated by off-axis
sources, potentially creating errors in the measured flux density of
the target source.  Direction-dependent calibration techniques are
still under investigation \citep{bhat}. Preliminary results from the
fields not affected by these effects show significant variability
(Goedhart~S., van~Rooyen~R. van~der~Walt~D.J.,Maswanganye~J.P., Gaylard~M.J. in preparation). The r.m.s. noise after continuum subtraction is typically 0.15 Jy per 381 Hz channel after 50 minutes of integration time. Observations centered on  \mbox{G330.89-0.36}, which had a
peak brightness of 824\,Jy\,per beam, were used to measure the spectral
dynamic range and channel isolation. The total time on source for this observation was 3.95 hours and an r.m.s. noise level of 75 mJy was achieved in line-free channels in the continuum-subtracted data cube. The line-to-line dynamic range
was 31\,dB. A maximum imaging dynamic range of 32\,dB was
achieved after applying self-calibration and \textsc{clean}ing
interactively.

\section{Summary}
\label{sec:lessons}

\noindent
We learned several lessons during the construction and use of the KAT-7 array. Here we enumerate the technological lessons learned and their impact on MeerKAT design, and summarize the unique science results captured during scientific commissioning.

\subsection{Technology Lessons Learned}

The lightweight Stirling coolers used for the cryogenic system are
cheaper than conventional Gifford-McMahon (G-M) cycle coolers but need
maintenance (at least) annually. They are not cold enough (80\,K) to
achieve a large cryopumping effect and so the system needs ion pumps
in order to retain a high vacuum and mechanical dampers to reduce the
vibration generated by the Stirling coolers. Both of these sub-systems
need replacement on a regular basis, which involves the use of cranes
and a team of technicians, plus about 2 days on an external pump to
obtain high enough vacuum before the receivers can be used.  The
capital expenditure for Stirling cycle cooling for 3 receivers on all
64 MeerKAT antennas is ZAR (South African Rand) 3.3 million less, but the cost of expected
maintenance, parts and electricity bill would be ZAR 1.1 million
more. This means that over a 3-4 year period Stirling cycle coolers
would end up much more expensive to install and operate than G-M
cryogenic coolers. Using the G-M cooling system should also reduce
down-time for MeerKAT antennas and this has not been factored into the
costs.

The composite dishes with embedded metal mesh have large weight
advantages over dishes made of large metal panels which in turn means
that the backing structure can be light and the motors need not be
powerful.  They must be constructed correctly as they cannot be
adjusted or machined after the fact, and this in turn means that
the fibre weave, temperature, humidity and vacuum need to be very
carefully controlled during fabrication.  Also, the mould needs to be
machined to higher accuracy than is needed for the dish; this is
relatively simple for a circularly symmetrical design, as only one
segment has to be accurately made and then copied.  This is far more
difficult for a design without that symmetry (e.g.~MeerKAT).  Large
single-piece dishes must be made on-site as moving them becomes a
logistical problem.

The ball-screw mechanism for the elevation drive has proven reliable
and simple and it is retained in the MeerKAT design.

The KATCP protocol has shown itself to be flexible and very useful in
testing and commissioning, and can readily be put into scripting form
for routine observations.

SPEAD, although still undergoing active development, has also proven
effective and will be further rolled out for use in MeerKAT.

The correlator architecture, correlator and beamformer based on FPGA
technology and data transfer over a managed network switch, have also
been found to be both flexible and reliable.

\subsection{Commissioning science results}

Through commissioning we have found that KAT-7 is well suited to observations addressing a diverse range of scientific interests.  The telescope's strengths include mapping extended sources in both continuum and spectral-line emission, monitoring variable sources, conducting high time0resolution observations.  In particular, the scientific results from KAT-7 commissioning include:

\begin{itemize}
\item mapping low-surface-brightness continuum emission (e.g.~radio
  halos and relics, and supernova remnants);
\item mapping extended spectral-line sources, such as HI around nearby
  galaxies, and the large scale structure around galaxy clusters;
\item high frequency resolution of OH maser emission around star-forming regions;
\item observing the variability of continuum sources $\geq 10$\,mJy at
  1.8\,GHz;
\item high time-resolution pulsar observations with the beamformer output;
\item VLBI observations with the beamformer output.
\end{itemize}

For continuum sources we reach a confusion limit of about
1\,mJy after 8\,hours (@1822 GHz), depending on the field. %, and this is worse in the Galactic plane.
The confusion limit makes high-dynamic-range imaging
difficult, although dynamic ranges of a few thousand are regularly
reached. For spectral-line observations, this confusion limit is not an
issue. Finally, KAT-7 has pioneered African VLBI observations in combination with HartRAO, in preparation for the SKA.

%% file: tex/acknowledgements.tex
\section{Acknowledgements}

\noindent
The SKA South Africa project was formed in 2004 and is funded by
the South African Department of Science and Technology (DST), and
administered by the National Research Foundation (NRF).

We would especially like to acknowledge the help of the Adam Deller
and Walter Brisken with running the DiFX correlator package. Jonathan
Quick helped with running the HartRAO VLBI system for the other end of
the VLBI test.

We would also like to thank the unnamed reviewer of this paper who
suggested many improvements.

%% file: tex/places.tex
\vspace{5mm}
\textit{
$^1$Square Kilometer Array South Africa, The Park, Park Road, Pinelands, Cape Town 7405, South Africa\\
$^2$Deptartment of Physics, University of Oxford, Parks Road, Oxford OX1 3PU, UK\\
$^3$EMSS Antennas, 18 Techno Avenue, Technopark, Stellenbosch 7600, South Africa\\
$^4$School of Physics, University of the Witwatersrand, PO BOX Wits, Johannesburg 2050, South Africa\\
$^5$Dept. of Astronomy, University of Cape Town, Private Bag X3, Rondebosch 7701 Cape Town, South Africa\\
$^6$Netherlands Institute for Radio Astronomy (ASTRON), Postbus 2, 7990 AA, Dwingeloo, The Netherlands\\
$^7$Kapteyn Astronomical Institute, University of Groningen, Postbus 800, 9700 AV Groningen, The Netherlands\\
$^8$Rhodes University, P.O. Box 94, Grahamstown 6140, South Africa\\
$^9$Dept. of Electrical Engineering, University of Cape Town,University Private Bag Rondebosch, 7701 Cape Town, South Africa\\
$^{10}$Hartebeesthoek Radio Astronomy Observatory, P.O.Box 443, Krugersdorp 1740South Africa\\
$^{11}$South African Astronomical Observatory, P.O. Box 9, Observatory 7935,South Africa\\
$^{12}$Department of Physics \& Astronomy, University of the Western Cape, Private Bag X17. Bellville 7535, Cape Town, South Africa\\
$^{13}$Station de Radioastronomie de Nan\c{c}ay, Observatoire de Paris, PSL Research University, CNRS, Univ. Orl\'{e}ans, OSUC, 18330 Nan\c{c}ay, France
$^{14}$National Radio Astronomy Observatory,1003 Lopezville Road, Socorro, New Mexico, 87801,\\
$^{15}$Fuller Theological Seminary, 135 N. Oakland Avenue, Pasadena, California, 91182. U.S.A\\
$^{16}$Skaha Remote Sensing Ltd, 3165 Juniper Dr, Naramata BC V0H 1N1, Canada\\
}